\documentclass[prb,twocolumn]{revtex4}    
\usepackage{graphics}
\begin{document}
                                                                                                                             
                                                                                                                             
\title{ Density functional theoretical (DFT) study for the prediction of spectroscopic parameters of ClCCCN.}                              
                                                                                                                             
\author{Pradeep Risikrishna Varadwaj}\email{pr.varadwaj@saha.ac.in}
                                                                                                                             
\affiliation{Saha Institute of Nuclear Physics, Block-AF, Bidhannagar, Kolkata - 700 064, India}

\begin{abstract}                          
DFT(B3LYP, B3PW91) level calculations in conjunction with three different of basis sets have been 
used to investigate the variations in the bond lengths, dipole moment and rotational constants, 
IR frequencies, IR intensities and rotational invariants of ClCCCN. 
The nuclear quadrupole constants of chlorine and nitrogen of ClCCCN 
have been calculated on the experimental r$_{s}$ structure as well as on the B3PW91/6-311++g(d,p) optimized 
geometry and are found to be within the scale length of the experimental uncertainty. 
The slopes and intercepts obtained from the regression analysis between the B3LYP/6-311++g(d,p) level calculated  and
experimental B$_{o}$ values of ClCCCN were used
to calculate the reasonable values of rotational constants of all the rare isotopic species of ClCCCN 
having standard deviation $\pm$0.048 MHz. 
All the spectroscopic parameters obtained from DFT calculation shows satisfactory agreement with the 
available experimental data. \\
\end{abstract}
\maketitle
\section{\bf Introduction}
Quantum chemical calculation and spectroscopic characterization of organic compounds, free radicals, radical anions etc. 
have found considerable amount of interest in recent years \cite{guide1,guide2,guide3,guide4}. 
Accurate and efficient calculation of spectroscopic 
constants for a wide range of molecular systems employs readily available methods and basis sets \cite{guide5,guide6}. 
The DFT method has been demonstrated to 
have great accuracy in reproducing the experimental values of quadrupole hyperfine coupling constants \cite{guide7}, 
molecular structural properties \cite{guide8}, rotational 
constants \cite{guide1,guide4}, IR frequencies, IR intensities and rotational invariants 
\cite{guide9,guide10,guide11} etc. to within 8-10$\%$. 
DFT studies have been carried out for many compounds like BrCCCN \cite{guide12}, 
FC$_8$H \cite{guide13}, ClCN \cite{guide14} etc. 
and benzene derivatives like o-benzyne \cite{guide4}, 3-chlorobenzonitrile \cite{guide15} etc. to calculate 
their accurate spectroscopic parameters. Highest number of the polyatomic molecules identified in space are mostly 
carbon chains \cite{guide16,guide17}, the longest chain identified so far is HC$_{11}$N \cite{guide18}. 
Identification and characterization of these species 
could have been made by their previously reported spectroscopic parameters. Spectroscopic studies of the 
related series of molecules containing halogens have been extensively studied \cite{guide19,guide20,guide21,guide22}. 
Recently, we have investigated the variations of the spectroscopic 
parameters of bromocyanoacetylene (BrCCCN) at the HF-SCF and DFT levels in conjunction with a 
variety of basis sets \cite{guide12}. 
It was found from the the investigation that DFT-B3LYP/6-311++g(d,p) level 
calculation is satisfactory in the prediction of bond lengths and rotational constants etc. of BrCCCN when compared 
against their 
experimental values. However, to my knowledge, a detailed quantum chemical calculations and molecular properties of 
chlorocyanoacetylene (ClCCCN) is very limited. Analysis of the microwave spectra 
of ClCCCN has been reported by T. Bjorvatten \cite{guide23}. 
Infrared studies of ClCCCN have been reported by S.J. Cyvin et.al \cite{guide24}. Later on, 
P. Klaboe et. al\cite{guide25,guide26} have reported the Raman spectral analysis and revised their 
vibrational assignments in the region 100-4000 cm$^{-1}$. \\
$~~~$In this study, the results of 
DFT (B3LYP, B3PW91) level calculations of optimized molecular geometry, rotational constants, 
quadrupole coupling constants, IR fundamental frequencies, IR intensities and rotational invariants of ClCCCN are 
compared against their available experimental data and reported. 
Linear regression analysis between the B3LYP/6-311++g(d,p) level 
calculated and experimental B$_{o}$ values have been made for a reasonable prediction of rotational constants of all the other 22 rare isotopomers of ClCCCN. 
Due to the lack of quantitative data on the IR intensities in the literature, 
the rotational invariants \cite{guide27,guide28,guide29} of ClCCCN have been predicted and discussed 
here may be helpful in future for the experimental IR spectroscopist in interpreting IR intensities of this molecule. 
Satisfactory agreements between the calculated and available experimental values 
of spectroscopic parameters 
have been found at the B3LYP level and compared with the corresponding values calculated at B3PW91, RHF and MP$_{2}$ 
levels of theory. In addition, 
a test of the optimization followed by a frequency calculation at the MP$_{2}$
level in conjunction with a medium size basis set 6-31G or/and 6-311++g(d,p) predicts two negative frequencies 
 (saddle point of oder two) corresponding to the -C$\equiv$C-Cl doubly degenerate bending mode ($\nu_{7}$) 
and thus it was difficult to calculate the harmonic frequencies of these modes under C$_{v}$ point group 
symmetry.
\section{Computational Methods and Calculations}
Geometry optimization and quantum chemical calculations
were carried out at the restricted HF-SCF method, MP$_{2}$ method and Density Functional Theoretical (DFT) 
method under C$_{v}$ point group symmetry.
The hybrid HF/DFT methods used were Becke's
three-parameter method \cite{guide30} with Lee-Yang-Parr correlation (B3LYP),
Becke's three parameter exchange with Perdew-Wang
Correlation \cite{guide31} (B3PW91) and Becke's
one-parameter method with Lee-Yang-Parr correlation (B1LYP) \cite{guide32,guide33}. Five different basis sets 
used during the analysis were Dunnings correlation consistent polarized valence 
double and triple zeta basis sets aug-cc-PVNZ (N $=$ D, T) 
\cite{guide34} augmented with a d function, 
Ahlriches TZV(3df,2p) \cite{guide35} and triple split zeta qualities 6-311++g(d,p) and 
6-311+g(df,pd) augmented with p, d and f functions. Tight convergence
in stead of default convergence criteria was used with ultrafine integration grid for the calculation of 
normal mode frequencies.
All calculations were carried out by the Gaussian03W suite program package \cite{guide36}. \\
$~~~$The physical quantities $\bar{p}_{\alpha}$, $\beta_{\alpha}$ and $\chi_{\alpha}$
are invariant with respect to the
rotation of space-fixed coordinate axes which are known as generalized atomic
polar tensor charge (GAPT) \cite{guide11}, atomic
anisotropy and Kings atomic effective charge of $\alpha$th atom respectively and are useful for interpreting 
infrared intensities. The components of the (3$\times$3) atomic polar tensor matrix $P^{(\alpha)}_{X}$ are
defined as the first derivatives of the components of molecular dipole moment with respect to the atomic
Cartesian displacement co-ordinates of each atom $\alpha$ ($=$Cl, C and N) of ClCCCN  given by: \\

\[P^{(\alpha)}_{X} = \left(\begin{array}{ccc}
\frac{\partial p_{x}}{\partial x_{\alpha}} & \frac{\partial p_{x}}{ \partial y_{\alpha}} & \frac{\partial p_{x}}{\partial z_{\alpha}} \\
\frac{\partial p_{y}}{\partial x_{\alpha}} & \frac{\partial p_{y}}{\partial y_{\alpha}} & \frac{\partial p_{y}}{\partial z_{\alpha}}\\
\frac{\partial p_{z}}{\partial x_{\alpha}} & \frac{\partial p_{z}}{\partial y_{\alpha}} & \frac{\partial p_{z}}{\partial z_{\alpha}}\\
\end{array} \right) = \left(\begin{array}{ccc}
{p_{xx}} & {p_{xy}} & {p_{xz}} \\
{p_{yx}} & {p_{yy}} & {p_{yz}} \\
{p_{zx}} & {p_{zy}} & {p_{zz}} \\
\end{array} \right) ....  (1)\] 

The generalized atomic polar tensor charges (GAPTs) are nothing but mean
dipole derivatives \cite{guide27,guide28} given by: \\
                                                                                                                             
${\bar p}_{\alpha} = \frac{1}{3}(\frac{\partial p_{x}}{\partial x_{\alpha}} +
\frac{\partial p_{y}}{\partial y_{\alpha}} + \frac{\partial p_{z}}{\partial z_{\alpha}})$ $~~~~~~~~~~~~~~~~~~~~~~~~~~~~~~~~~~~~~~(2)$\\
                                                                                                                             
such that \\
                                                                                                                             
$\sum_{\alpha} {\bar p_{\alpha}} = 0$ $~~~~~~~~~~~~~~~~~~~~~~~~~~~~~~~~~~~~~~~~~~~~~~~~~~~~~~~~~~~~~~~~~~~~~~~~~~ (3)$ \\
                                                                                                                             
for the neutral molecule ClCCCN. The quantity $\chi_{\alpha}$ known as the effective charge of
the $\alpha$th atom, defined by King and co-workers \cite{guide28,guide29}
as: \\

$\chi^{2}_{\alpha} = \frac{1}{3} [Tr.(P^{(\alpha)}_{X}P'^{(\alpha)}_{X})]$ = $\bar{p}^{2}_{\alpha} + \frac{2}{9} \beta^{2}_{\alpha}$ $~~~~~~~~~~~~~~~~~~~~~~~~~(4)$\\
                                                                                                                             
where \\
                                                                                                                             
$\beta^{2}_{\alpha} = \frac{1}{2}[(p^{(\alpha)}_{xx}-p^{(\alpha)}_{yy})^{2} + (p^{(\alpha)}_{yy}-p^{(\alpha)}_{zz})^{2} + (p^{(\alpha)}_{zz}-p^{(\alpha)}_{xx})^{2}] + \frac{3}{2} (p^{(\alpha)^{2}}_{xy} + p^{(\alpha)^{2}}_{yz} + p^{(\alpha)^{2}}_{xz} + p^{(\alpha)^{2}}_{zx} + p^{(\alpha)^{2}}_{yx} + p^{(\alpha)^{2}}_{zy})$ $~~~~~~~~~~~~(5)$ \\
                                                                                                                             
The molecular polar tensor of ClCCCN is a juxtaposition of the (3$\times$3) atomic polar tensors given by: \\
                                                                                                                             
$P_{X} = \{P^{Cl}_{X}|P^{C}_{X}|P^{C}_{X}|P^{C}_{X}|P^{N}_{X}\}$ $~~~~~~~~~~~~~~~~~~~~~~~~~~~~~~~~~~~(6)$\\
                                                                                                                             
and is calculated from \\
                                                                                                                             
$P_{X} = P_{Q}L^{-1}UB + P_{\rho}\beta$ $~~~~~~~~~~~~~~~~~~~~~~~~~~~~~~~~~~~~~~~~~~~~~~~~~~~~~~~~~~~~(7)$\\
                                                                                                                             
where $L^{-1}$, U and B are matrices \cite{guide37} used for molecular vibrational analysis and
$P_{Q}$ contains the dipole moment derivatives
with respect to the normal co-ordinates, which are proportional to
the experimental infrared intensities \cite{guide38}. $P_{\rho}$ is the rotation plus translation polar
tensor whose elements for a neutral molecule proportional to
$\mu/I^{\frac{1}{2}}$ and $\rho = BX$, where $\mu$ and I are dipole moment and moment of inertia of of the molecule 
respectively. Atomic polar tensor components were directly calculated by Gaussian03 package.\\
\begin{table*}[htb]
\caption{Comparison of the molecular optimized geometry (in \AA), dipole moment, rotational constant and total energy
(a.u.) of ClCCCN (calculated by various methods in conjunction with basis sets of increasing size) with the 
experimental average r$_{s}$ values \cite{guide23}.}
\label{table:I}
\newcommand{\m}{\hphantom{$-$}}
\newcommand{\cc}[1]{\multicolumn{1}{c}{#1}}
\renewcommand{\tabcolsep}{0.15pc} 
\renewcommand{\arraystretch}{1.2} 
\begin{tabular}{@{}ccccccccccccc}
\hline
& \cc{Method}&\cc{basis} &\cc{Cl-C} &\cc{C$\equiv$C} &\cc{C-C} &\cc{C$\equiv$N} &\cc{Cl..N}& \cc{$\mu$/D} & \cc{B/MHz} &\cc{$^{^{35}Cl}\chi_{aa}$} &\cc{$^{N}\chi_{aa}$} & \cc{Energy}\\
\hline
&HF          &6-311++g(d,p) &\m1.6359  &\m1.1809 &\m1.3855 &\m1.1355 &\m5.3378 &\m3.82 &\m1394.78  &\m-81.89&\m-3.81 & \m-627.5026\\
&            &aug-cc-pVDZ   &\m1.6421  &\m1.1881 &\m1.3914 &\m1.1372 &\m5.3588 &\m3.85 &\m1381.38  &\m-80.34&\m-3.61 &\m-627.4784  \\&            &aug-cc-PVTZ   &\m1.6325  &\m1.1780 &\m1.3852 &\m1.1277 &\m5.3234 &\m3.86 &\m1399.53  &\m-82.34  &\m-4.30 &\m-627.5349\\
&MP$_{2}$    &6-311++g(d,p) &\m1.6294  &\m1.2238 &\m1.3705 &\m1.1826 &\m5.4063 &\m4.12 &\m1362.79  &        &        & \m-628.1965\\
&B3LYP       &6-311++g(d,p) &\m1.6322  &\m1.2064 &\m1.3647 &\m1.1592 &\m5.3625 &\m4.17 &\m1382.60  &\m-72.82&\m-3.48 & \m-629.2369\\
&            &aug-cc-pVDZ   &\m1.6402  &\m1.2159 &\m1.3721 &\m1.1670 &\m5.3952 &\m4.17 &\m1366.05  &\m-77.31  &\m-2.10  & \m-629.2044\\
&            &aug-cc-pVTZ   &\m1.6276  &\m1.2039 &\m1.3643 &\m1.1562 &\m5.3520 &\m4.14 &\m1380.00 &\m-76.10 &\m-3.64  &\m-629.2646\\
&B3PW91      &6-311++g(d,p) &\m1.6232  &\m1.2073 &\m1.3632 &\m1.1595 &\m5.3532 &\m4.19 &\m1387.85  &\m-74.93 &\m-3.38  & \m-629.1045\\
&            &aug-cc-pVDZ   &\m1.6312  &\m1.2160 &\m1.3700 &\m1.1671 &\m5.3843 &\m4.13 &\m1371.95  &\m-76.05  &\m-2.84   & \m-629.0762\\
&            &aug-cc-pVTZ   &\m1.6196  &\m1.2049 &\m1.3630 &\m1.1567 &\m5.3442 &\m4.16 &\m1392.34&\m-74.16  &\m-3.54   & \m-629.1332 \\
&Expt. &              &\m1.6245  &\m1.2090 &\m1.3690 &\m1.1602 &\m5.3627&\m3.38    &\m1382.328(2)&         &          & \\
\hline
\end{tabular}\\[2pt]
\end{table*}
\begin{table*}[htb]
\caption{Comparison of calculated$^{a}$ and experimental\cite{guide23} values of quadrupole coupling constants $\chi_{aa} (=\chi_{zz})$ (in MHz) of 
$^{35}$Cl, $^{37}$Cl and $^{14}$N of ClCCCN. The values given in the parentheses are directly 
calculated without using the scaling factors \cite{guide7}.}
\label{table:II}
\newcommand{\m}{\hphantom{$-$}}
\newcommand{\cc}[1]{\multicolumn{1}{c}{#1}}
\renewcommand{\tabcolsep}{1.8pc} 
\renewcommand{\arraystretch}{1.3} 
\begin{tabular}{@{}cccccccccccc}
\hline
&\cc{Atoms}&\cc{$\chi^{b}_{r_{s}}$}  &\cc{$\chi_{B3LYP}$}  &\cc{$\chi_{B3PW91}$} &\cc{$\chi_{Expt.}$} \\
\hline
& $^{35}$Cl          &\m-79.57(-80.39)   &\m-80.67(-81.48) &\m-79.53(-80.35)   &\m-75$\pm$4 \\
& $^{37}$Cl          &\m-62.94(-63.50)   &\m-63.58(64.14)  &\m-62.68(-63.25)   &\m-62$\pm$3 \\
&$^{14}$N($^{35}$Cl)&\m-4.29            &\m-4.30          &\m-4.31             &\m...       \\
& $^{14}$N($^{37}$Cl)&\m-4.30            &\m...             &\m...              &\m...       \\
\hline
\end{tabular}\\[2pt]
$^{a}$ See text for discussion. \\
$^{b}$ Ref.\cite{guide23}. Actual bond distances based on the r$_{s}$ co-ordinates are reported to be 1.6233(31) (1.6256(32)), 1.2086(42) (1.2093(43)), 1.3700(23) (1.3680(21)), 1.1607(10) (1.1596(10)) for Cl-C, C$\equiv$C, C-C, C$\equiv$N of 35 (and 37) species of ClCCCN respectively. \\
\end{table*}
\begin{table}[htb]
\caption{Comparison of the rotational constants of all the isotopomers of ClCCCN calculated at 
B3LYP/6-311++g(d,p) level against their experimental values \cite{guide23}.}
\label{table:III}
\newcommand{\m}{\hphantom{$-$}}
\newcommand{\cc}[1]{\multicolumn{1}{c}{#1}}
\renewcommand{\tabcolsep}{.25pc} 
\renewcommand{\arraystretch}{1.2} 
\begin{tabular}{@{}cccccc}
\hline
& \cc{Isotopic Species} &\cc{Calc.} & \cc{Scaled$^{a}$} &\cc{Expt.} \\
\hline
&$^{35}ClCCCN$      			&\m1382.602   &\m1382.339         &\m1382.328(2)      \\
&$^{35}Cl^{13}CCCN$ 			&\m1381.751   &\m1381.488           &\m1381.454(2)      \\
&$^{35}ClC^{13}CCN$ 			&\m1380.607   &\m1380.346           &\m1380.357(2)     \\
&$^{35}ClCC^{13}CN$ 			&\m1366.338   &\m1366.090           &\m1366.054(2)     \\
&$^{35}ClCCC^{15}N$ 			&\m1344.232   &\m1344.005          &\m1343.913(2)     \\
&$^{35}Cl^{13}CCC^{15}N$ 		&\m1343.295   &\m1343.068 &        \\
&$^{35}ClC^{13}CC^{15}N$ 		&\m1342.535   &\m1342.309 &       \\
&$^{35}ClCC^{13}C^{15}N$ 		&\m1329.395   &\m1329.181 &       \\
&$^{35}Cl^{13}C^{13}CCN$ 		&\m1379.727   &\m1379.466 &       \\
&$^{35}Cl^{13}C^{13}CC^{15}N$ 		&\m1341.571   &\m1341.346 &       \\
&$^{35}Cl^{13}CC^{13}CN$ 		&\m1365.420   &\m1365.173 &       \\
&$^{35}Cl^{13}CC^{13}C^{15}N$ 		&\m1328.391   &\m1328.178 &       \\
&$^{35}ClC^{13}C^{13}CN$ 		&\m1364.517   &\m1364.271 &       \\
&$^{35}ClC^{13}C^{13}C^{15}N$ 		&\m1327.846   &\m1327.634 &       \\
&$^{35}Cl^{13}C^{13}C^{13}CN$ 		&\m1363.571   &\m1363.325 &       \\
&$^{35}Cl^{13}C^{13}C^{13}C^{15}N$ 	&\m1326.817   &\m1326.790 &      \\
\\
&$^{37}ClCCCN$ 				&\m1350.544   &\m1350.311            &\m1350.360(2)     \\
&$^{37}Cl^{13}CCCN$ 			&\m1349.888   &\m1349.655            &\m1349.692(2)      \\
&$^{37}ClC^{13}CCN$ 			&\m1348.379   &\m1348.148            &\m1348.202(2)     \\
&$^{37}ClCC^{13}CN$ 			&\m1334.300   &\m1334.082            &\m1334.089(2)     \\
&$^{37}ClCCC^{15}N$ 			&\m1312.831   &\m1312.633            &\m1312.603(2)     \\
&$^{37}Cl^{13}CCC^{15}N$        	&\m1312.099   &\m1311.901  &       \\
&$^{37}ClC^{13}CC^{15}N$ 		&\m1310.977   &\m1310.780  &      \\
&$^{37}ClCC^{13}C^{15}N$        	&\m1298.001   &\m1297.827  &      \\
&$^{37}Cl^{13}C^{13}CCN$        	&\m1347.697   &\m1347.466  &      \\
&$^{37}Cl^{13}C^{13}CC^{15}N$   	&\m1310.220   &\m1310.024  &      \\
&$^{37}Cl^{13}CC^{13}CN$        	&\m1333.585   &\m1333.367  &      \\
&$^{37}Cl^{13}CC^{13}C^{15}N$   	&\m1297.209   &\m1297.025  &      \\
&$^{37}ClC^{13}C^{13}CN$        	&\m1332.317   &\m1332.108  &      \\
&$^{37}ClC^{13}C^{13}C^{15}N$   	&\m1296.302   &\m1296.119  &      \\
&$^{37}Cl^{13}C^{13}C^{13}CN$   	&\m1331.576   &\m1331.360  &      \\
&$^{37}Cl^{13}C^{13}C^{13}C^{15}N$      &\m1295.486   &\m1295.304  &      \\
\hline
\end{tabular}\\[2pt]
$^{a}$ Scaled using the slopes and intercepts obtained from the linear regression analysis.
\end{table}
$~~~$On the other hand, nuclear quadrupole coupling constants (NQCCs) of chlorine and nitrogen of ClCCCN 
are calculated by using the electric field gradients (EFGs). The elements of the NQCC tensors
$\chi_{ij}$ are related to those of the EFG tensors $q_{ij}$ by : \\
                                                                                                               
$\chi_{ij} = (eQ/h) q_{ij}$ $~~~~~~~~~~~~~~~~~~~~~~~~~~~~~~~~~~~~~~~~~~~~~~~~~~~~~~~~~~~~~~~~~~~~~~~(8)$\\
                                                                                                               
where e is the fundamental electronic charge, h is Plank's constant
and $i,j = a, b, c$ are the principal axes of the inertia tensor. \\
\section{Results and Discussion}
Optimization of geometry for each molecule at the restricted HF-SCF and DFT levels 
were performed in the ground state by single point 
energy calculation. In each case, stationary points were found and 
linear geometry of ClCCCN was confirmed. Finally, a frequency calculation following 
each optimization have been performed in order to check for the existence of a true minimum and 
to confirm an equilibrium structure. 
A test of geometry optimization followed by a frequency calculation under C$_{v}$ 
point group symmetry at the MP$_{2}$ 
level in conjunction with 6-31g or/and 6-311++g(d,p) basis sets succeeded in locating a stationary point but 
with two imaginary 
frequencies (IMAG $=$ 2) corresponding to the -C$\equiv$C-Cl doubly degenerate bending mode. 
This confirms that the stationary point located by the MP$_{2}$ level of theory is not a true 
minimum rather a saddle point of order two on the potential energy surface of ClCCCN which may cause a large 
deformation to the geometry of the molecule. 
Similar features have also been reported elsewhere \cite{guide39}. 
So the scan of the potential energy surface should be 
emphasized in order to characterize the the nature of the saddle point on the 
potential energy surface of ClCCCN and to check any possible existance of conical intersection between the ground and 
first excited state. However, the DFT and HF-SCF level calculations were not able predict any imaginary 
frequencies implying that the stationary point is located at the global minimum of the potential energy hyper-surface. 
The results of our 
calculation for bond lengths of ClCCCN are summarized in Table 1. The 
experimental $r_{s}$-structural values \cite{guide23} are also given for comparison. 
As can be seen from Table 1, the HF-SCF level calculations 
highly overestimates Cl-C and C-C bond lengths and underestimates 
C$\equiv$C and C$\equiv$N bond lengths, whereas these 
values calculated at the MP2 level overestimates all these bond lengths when compared against their 
experimental values. 
However, the results of DFT-B3LYP, B3PW91 method in conjunction with 6-311++g(d,p) 
basis set are in rather pleasing agreement with the experimental $r_{s}$ values. Here, We note a slight 
overestimation of Cl-C bond length and an
underestimation of C$\equiv$C and C$\equiv$N bond lengths, differences between calculated values
and the average $r_{s}$ values in bond lengths for ClCCCN are 0.0077 \AA, -0.0026 \AA, -0.0043 \AA$~$ and -0.001 \AA$~$ for 
Cl-C, C$\equiv$C, C-C and C$\equiv$N respectively at the B3LYP level, whereas a slight underestimation of all these 
bond distances being noticed at the B3PW91 level of theory, differences are -0.0013 \AA, -0.0017 \AA, -0.0058 \AA, 
-0.0007 \AA$~$ for Cl-C, C$\equiv$C, C-C and C$\equiv$N respectively. 
A comparison of bond lengths calculated at various levels, summarized in Table 1, indicate 
that convergence has not achieved on improving the size of 
the basis sets 6-311++g(d,p) to aug-cc-PVTZ. 
However, a better comparison is made on the over-all (end-over-end) length , $r_{Cl...N}$, of the molecule which is
much less sensitive to the experimental and theoretical uncertainty. The average value of this length
at the B3LYP level in conjunction with 6-311++g(d,p) basis set 
is calculated to be 5.3625 \AA$~$ which differs from the average experimental value by 
a factor of 0.0002 \AA$~$. Thus, it is to say that the over-all 
$r_{s}$-structure is a good approximation to the over-all bond length at the correlated
level B3LYP/6-311++g(d,p) within, say, 0.001 \AA$~$, which is mostly under the statistical uncertainties. \\
\begin{figure}
\resizebox{9.0cm}{2.2cm}
{\includegraphics{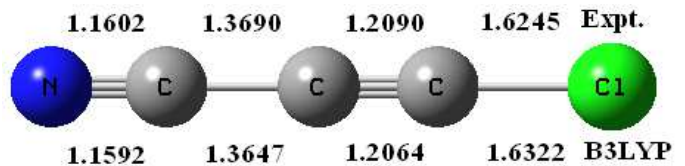}}    
\caption{Comparison between the calculated (B3LYP/6-311++g(d,p)) and experimental overall geometry of ClCCCN.}
\label{fig1}                                 
\end{figure}
\begin{figure}
\resizebox{7.5cm}{8.5cm}
{\includegraphics{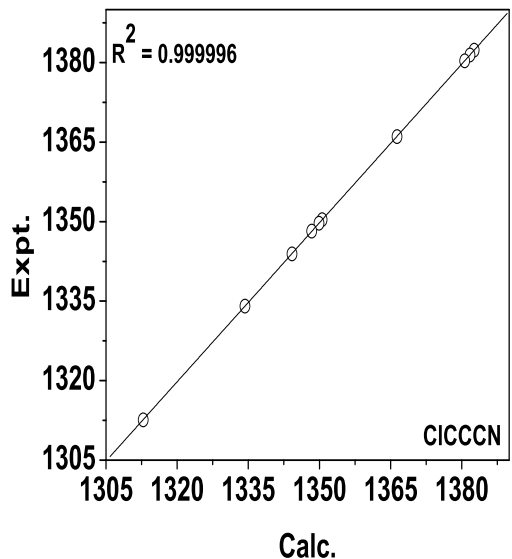}}    
\caption{Comparison between the calculated (B3LYP/6-311++g(d,p)) and experimental rotational constants of ClCCCN. 
The slopes and intercepts obtained from the linear regression analysis are 0.99907 and 1.02283 respectively.}
\label{fig2}                                 
\end{figure}
$~~~$Interestingly, our B3LYP/6-311++g(d,p) level optimized geometry leads to 
satisfactory values of rotational constants of ClCCCN shown in Table 3. 
The experimental values are also given for comparison. 
Rotational constants of ClCCCN calculated at B3LYP/6-311++g(d,p) level are a little higher (0.02 $\%$) than the 
experimental $B_{o}$ values of all the 10 different isotopomers, whereas these values calculated in 
conjunction with other basis sets are far off (summarized in Table 1). 
Thus, the agreement between the theoretical prediction and experimental 
observation of rotational constants clearly indicates that the B3LYP/6-311++g(d,p) 
optimized average geometry, shown in Fig 1, is realistic and advocates for a decent 
predictive power of the computational procedure applied to ClCCCN, though differences are 
more than the error limits for the $r_{e}$ structures. Based on the linear regression analysis, shown in Figures 2, 
between the calculated and 
experimental B$_{o}$ values, rotational constants of other 22 rare isotopomers of ClCCCN has been 
predicted accurately having standard deviation $\pm0.048$ MHz. A closer look at the 
Table 1 indicates that the HF-SCF wave functions predict dipole moments of ClCCCN 
close to its experimental value, deviation being 0.4 D where as the DFT wave functions overestimates it by a factor of 
0.72 D irrespective of the basis sets used. Zero point vibrational effects are
not taken into consideration in all these calculations. \\
$~~~$The nuclear quadrupole coupling
constants (NQCCs) of $^{35}$Cl, $^{37}$Cl and $^{14}$N of ClCCCN, NQCC
being proportional to the the electric field gradient (EFG), are calculated by using equation (8). 
Calculation of EFGs being made on the $r_{s}$-structures \cite{guide23} as well as 
on the B3LYP and B3PW91/6-311++g(d,p) optimized geometries of ClCCCN. 
The methods, basis sets and procedures \cite{guide7} used during the 
calculation were B3PW91/6-311+g(df,pd) for nitrogen and B1LYP/TZV(3df,2p) for chlorine respectively. 
The coefficients $(eQ/h)_{eff.}$ in equation (8), 
obtained from the linear regression analysis between the calculated EFGs and experimental 
NQCCs for a series of molecules containing chlorine and nitrogen, 
are reported \cite{guide7} to be 4.5586(40) MHz/a.u., -19.185 MHz/a.u. and -15.120 MHz/a.u. 
for $^{14}$N, $^{35}$Cl and $^{37}$Cl respectively. 
The nuclear quadrupole coupling constants of  $^{35}$Cl, $^{37}$Cl and $^{14}$N are calculated on the $r_{s}$ structure 
is summarized in Table 2. The values of NQCCs calculated on the B3LYP, B3PW91/6-311++g(d,p) 
optimized geometries are also included 
in Table 2 for comparison. As can be seen, the values of NQCCs deviates from 
the experimental values by a factor of 4.57 MHz (6.09 $\%$) and 0.94 MHz (1.5 $\%$) for $^{35}$Cl and $^{37}$Cl 
respectively. These deviations are significantly due to the largest error $\pm$4 MHz and $\pm$3 MHz present in the 
measurement of the quadrupole hyperfine structures of $^{35}$Cl and $^{37}$Cl of ClCCCN, though the calculated values 
are within the scale length of experimental uncertainties. A comparison of NQCCs of Chlorine 
calculated on the B3PW91/6-311++g(d,p) 
geometry as well as on the r$_{s}$ structure reveals an almost similar value, which is basically due to the 
similar C-Cl bond length, whereas its value calculated on the B3LYP/6-311++g(d,p) optimized 
geometry is a bit off as expected since NQCC varies linearly with 
C-Cl bond length. A comparison with ClCN \cite{guide14} shows an 
underestimation in the value of NQCC of Cl, 
which being supported by a decrease of C-Cl bond length, the decrease being within 5$\%$. 
On the other hand, the nuclear quadrupole coupling constant of $^{14}$N of ClCCCN is 
also summarized in Table 2. As can be seen, the values of NQCCs of $^{14}$N are very similar to each other 
since the r$_{s}$ structure as well as the optimized B3LYP, B3PW91/6-311++g(d,p) 
C$\equiv$N structures are predicted to be very similar. Finally, comparing 
FCCCN \cite{guide7}, HCCCN and DCCCN \cite{guide40} with that of ClCCCN 
reveals a similar C$\equiv$N structure resulting a similar value of NQCC. Thus, the 
accurate values of NQCCs of $^{35}$Cl, $^{37}$Cl and $^{14}$N are those derived from the r$_{s}$ 
structures.  Variations of NQCCs of $^{35}$Cl and N of ClCCCN calculated directly by various levels 
are also summarized in Table 1.\\   

$~~~$ ClCCCN has, in principle, ten fundamental modes. These modes under the C$_{v}$ point group symmetry
are distributed among four stretching vibrations of $\Sigma^{+}$ species and six bending 
vibrations of $\Pi$ species, out of which three $\Pi$ species are doubly degenerate.
Thus, the vibrational assignments of ClCCCN reported by
the co-authors \cite{guide24,guide25,guide26} comprises seven fundamental modes
$\nu_{1}$ , $\nu_{2}$, $\nu_{3}$, $\nu_{4}$, $\nu_{5}$, $\nu_{6}$ and $\nu_{7}$ which is in consistent
with the HF and DFT level calculations. All the normal modes are calculated to be Raman active. 
Harmonic frequencies have been calculated at the RHF
and DFT levels in conjunction with three different 
basis sets with increasing size in solution (benzene) as well as in vapour phase are 
summarized in Table 5. Experimental values are also 
included for comparison. All the predicted vibrational spectra have no imaginary frequency, implying that the
optimized geometry is locating at the global minimum of the potential energy
hyper-surface for both the methods. 
\begin{table*}[htb]
\caption{RHF, B3LYP and B3PW91 level (in conjunction with 6-311++g(d,p) (A), aug-cc-PVDZ (B), aug-cc-PVTZ (C)) 
calculated atomic mean dipole derivatives ($\bar {p}_{\alpha}$), 
tensor anisotropies ($\beta_{\alpha}$), charge
undeformability ($\bar{p}_{\alpha}/\beta_{\alpha}$) and effective charges ($\chi_{\alpha}$), Mullikan Charges 
($^{M}q_{\alpha}$) of $Cl_{1}C_{2}C_{3}C_{4}N_{5}$ in the unit of e.} 
\renewcommand{\arraystretch}{1.2} 
\renewcommand{\tabcolsep}{0.65pc} 
\label{table:VI}
\newcommand{\m}{\hphantom{$-$}}
\newcommand{\cc}[1]{\multicolumn{1}{c}{#1}}
\begin{tabular}{@{}cccccccccccc}
\hline\hline
& \cc{Invariants} & &\cc{RHF} & & &\cc{B3LYP}& & &\cc{B3PW91} &\\
&           &A&       B &C&A&         B&C&A&B           &C \\
\hline
&$\bar{p}_{Cl}$           &\m-0.144 &\m-0.128 &\m-0.126      & \m-0.158 &\m-0.148 &\m-0.145 &\m-0.165  &\m-0.154 &\m-0.150 \\
&$\bar{p}_{C}$            &\m0.398  &\m0.383 &\m0.385        & \m0.398  &\m0.394 &\m0.395  &\m0.410    &\m0.4021 &\m0.403\\
&$\bar{p}_{C}$            &\m-0.259 &\m-0.266 &\m-0.276      & \m-0.228 &\m-0.243 &\m-0.252 &\m-0.231  &\m-0.247 &\m-0.257 \\
&$\bar{p}_{C}$            &\m0.395  &\m0.408 &\m0.418        & \m0.341  &\m0.345 &\m0.359  &\m0.333    &\m0.342 &\m0.354\\
&$\bar{p}_{N}$            &\m-0.390 &\m-0.397 &\m-0.401      & \m-0.353 &\m-0.348 &\m-0.357 &\m-0.347  &\m-0.340 &\m-0.350\\
&$\beta_{Cl}$             &\m0.753 &\m0.725 &\m0.725         & \m0.926  &\m0.873 &\m0.864  &\m0.956    &\m0.896  &\m0.886\\
&$\beta_{C}$              &\m1.246 &\m1.229 &\m1.231         & \m1.430  &\m1.338 &\m1.329  &\m1.465   &\m1.361 &\m1.352\\
&$\beta_{C}$              &\m0.835 &\m0.813 &\m0.816         & \m0.890  &\m0.817 &\m0.809  &\m0.903    &\m0.819 &\m0.814\\
&$\beta_{C}$              &\m0.556 &\m0.487 &\m0.478         & \m0.558  &\m0.490 &\m0.475  &\m0.562    &\m0.483 &\m0.467\\
&$\beta_{N}$              &\m0.214 &\m0.179 &\m0.168         & \m0.173  &\m0.139 &\m0.131  &\m0.168    &\m0.129 &\m0.119\\
&$\bar{p}_{Cl}/\beta$     &\m-0.191&\m-0.176 &\m-0.174       & \m-0.171 &\m-0.169 &\m-0.168 &\m-0.173  &\m-0.172 &\m-0.169\\
&$\bar{p}_{C}/\beta$      &\m0.319  &\m0.312 &\m0.313        & \m0.278  &\m0.294 &\m0.297  &\m0.280    &\m0.295 &\m0.298\\
&$\bar{p}_{C}/\beta$      &\m-0.310 &\m-0.327 &\m-0.338      & \m-0.256 &\m-0.297 &\m-0.311 &\m-0.256  &\m-0.302 &\m-0.316\\
&$\bar{p}_{C}/\beta$      &\m0.710  &\m0.838 &\m0.874        &\m0.611  &\m0.704 &\m0.756  &\m0.592    &\m0.708 &\m0.758\\
&$\bar{p}_{N}/\beta$      &\m-1.822 &\m-2.218 &\m-2.387      & \m-2.040 &\m-2.504 &\m-2.725 &\m-2.005  &-2.636 &\m-2.941\\
&$|\chi_{Cl}|$            &\m0.383  &\m0.365 &\m0.364        & \m0.464  &\m0.437 &\m0.432  &\m0.480    &\m0.449 &\m0.444\\
&$|\chi_{C}|$             &\m0.710  &\m0.694 &\m0.697        & \m0.783  &\m0.744 &\m0.741  &\m0.804    &\m0.757 &\m0.754\\
&$|\chi_{C}|$             &\m0.471  &\m\m0.467 &\m0.473      & \m0.477  &\m0.455 &\m0.457  &\m0.483    &\m0.458 &\m0.462\\
&$|\chi_{C}|$             &\m0.474  &\m0.469 &\m0.474        & \m0.430  &\m0.416 &\m0.423  &\m0.425    &\m0.410 &\m0.417\\
&$|\chi_{N}|$             &\m0.403  &\m0.406 &\m0.409        & \m0.362  &\m0.354 &\m0.362  &\m0.355    &\m0.348 &\m0.354\\
&$^{M}q_{Cl}$		  &\m0.387  &-0.208 &\m-0.063        & \m0.463  &\m-0.030 &\m0.042  &\m0.510    &\m-0.137 &\m-0.003\\
&$^{M}q_{C}$		  &\m0.082  &-0.898 &\m-0.901        & \m 0.035 &\m-0.661 &\m-0.877 &\m0.101  &\m-0.678 &-0.820\\
&$^{M}q_{C}$		  &\m1.954  &\m1.838 &\m1.985        & \m1.648  &\m1.155 &\m1.656  &\m1.749    &\m1.367 &\m1.529\\
&$^{M}q_{C}$		  &\m-2.139 &\m-0.387 &\m-0.632      & \m-1.969 &\m-0.214 &\m-0.447 &\m-2.154  &\m-0.214 &\m-0.252\\
&$^{M}q_{N}$		  &\m-0.283 &\m-0.344 &\m-0.389      & \m-0.177 &\m-0.249 &\m-0.374 &\m-0.207  &\m-0.338 &\m-0.454\\
\hline\hline
\end{tabular}\\[2pt]
\end{table*}
Frequencies calculated at the
Hartree-Fock level contain known systematic errors due to the lack of electron correlation
and the choice of the basis sets used, resulting an overestimation of
10-12 $\%$. Therefore, it is usual to scale stretching frequencies predicted at the HF-SCF level by an emperical factor
0.893, which has been demonstrated to reproduce experimental values to a high degree accuracy for a wide
range of systems \cite{guide41}. However,  the improvement of the quality of the unscaled normal mode frequencies 
is significant in going from the 
non-correlated to the correlated level of theory and enlarging the size of the basis sets 
at the cost of its computational expanse. A comparison of the experimental values against the resulting 
IR normal mode frequencies calculated at the B3LYP level leads to a largest over-all error with 
6-311++g(d,p) basis set is 32 cm$^{-1}$, with 
the aug-cc-pVDZ basis set the over-all error is 16 cm$^{-1}$ whereas with aug-cc-pVTZ the over-all error is 18 cm$^{-1}$. 
An unexpected change of the 
$\nu_{6}$ fundamental frequency has been noticed both at the B3LYP and B3PW91 levels of calculation 
when 6-311++g(d,p) results being compared against the aug-cc-pVNZ (N $=$ D, T) results 
indicating that this mode is sensitive to the size of the basis sets used. Thus, the size of the aug-cc-pVDZ basis set 
is somewhat more precise for the prediction of normal mode frequencies of ClCCCN at the DFT levels of theory. 
The values of $\nu_{1}$,
$\nu_{2}$, $\nu_{3}$, $\nu_{4}$ and $\nu_{5}$ computed at B3LYP/aug-cc-pVDZ are overestimated from their
respective experimental values
by 85.1 $cm^{-1}$, 49.8 $cm^{-1}$, 25.1 $cm^{-1}$, 13.7 $cm^{-1}$, 13.2 $cm^{-1}$, 
whereas  $\nu_{6}$ and $\nu_{7}$ are underestimated by a factor of 21.4 $cm^{-1}$ and 7.4 cm$^{-1}$. 
These discrepancies can be corrected either by computing
an-harmonic force constants or introducing a scaled field or directly
scaling the calculated wave-numbers by taking the ratio between the calculated and observed frequencies for a 
particular type of motion. B3LYP scaling factor for stretching modes are all
close to 0.965 published elsewhere \cite{guide42}. Similarly, the normal mode frequencies calculated 
at the DFT-B3PW91 level of theory slightly overestimated from the DFT-B3LYP values, differences, for example, being 
2.9, 2, 1.3, 8.8, 
15.6, 9.3 and 13.6 cm$^{-1}$ for $\nu_{7}$...and $\nu_{1}$ respectively in conjunction with aug-cc-pVDZ basis set,
whereas this overestimation is highly overestimated for stretching and bending modes computed at the HF level of theory, 
differences being 20.9, 86.2, 84.7, 23.6, 36.2, 235.1 and 269.7 cm$^{-1}$ respectively. A comparison of the normal mode 
frequencies calculated at the B3LYP/aug-cc-pVDZ in solution (benzene) with those of gas phase values indicates a shift of 
C$\equiv$N stretching frequency towards the high wavelength region as expected whereas for other modes 
the shift is rather small 
which being in good agreement with the experimental shifts. On the other hand, a frequency calculation at the MP$_{2}$
level in conjunction with a medium size basis set 6-311++g(d,p), presented in Table 5, predicts two negative 
frequencies (saddle point of oder two) corresponding to the -C$\equiv$C-Cl doubly degenerate bending mode ($\nu_{7}$) and 
thus it was difficult to calculate the harmonic frequencies of these modes at this level.\\

$~~~$The infrared band intensities $I_{i}$, on the other hand, of ClCCCN corresponding to the seven vibrational modes
$\nu_{1}$, $\nu_{2}$, $\nu_{3}$, $\nu_{4}$,
$\nu_{5}$, $\nu_{6}$ and $\nu_{7}$ were assigned experimentally in gas phase as well as in solution (benzene) 
\cite{guide25,guide26} to be very strong,
strong, very week, weak, strong,
medium strong, and very weak respectively. The calculated intensities are summarized in Table 7. As can be seen, 
the most intense absorption infrared band (I$_{1}$) corresponds to the C$\equiv$N normal mode as expected irrespective of 
the methods and basis sets used, variation in intensity
is within 30 km/mole between the two (HF-SCF and DFT) theoretical methods. For the band
$I_{2}$, the differences between the RHF and B3LYP values in intensity are within 20 km/mole. 
Out of the three doubly degenerate bending modes, intensities of $I_{6}$ and $I_{5}$
corresponding to the normal modes $\nu_{5}$ and $\nu_{6}$
are calculated to be of equal in magnitude which were experimentally
assigned to be medium strong and strong respectively. 
As shown in Table 6, the main discrepancy in intensity is 
found for the $\equiv$C-C$\equiv$ stretching band $I_{3}$. For this band, variations in intensities 
are within 6-13 km/mole between the two theoretical methods.
A comparison of IR intensities calculated (for example, at the B3LYP/aug-cc-pVDZ level of theory) 
in benzene and those in vapour phase indicates a significant enhancement of 
C$\equiv$N stretching intensity as expected, since intensities measured in solutions somewhat larger 
than the gas phase intensities \cite{guide10}, whereas for other bands the change is very less. 
Thus, the agreement between the calculated and 
the experimental gauss treatment in intensities, is far from being quantitative. Since there is no
quantitative experimental data on IR intensity measurements for all these bands, no satisfactory interpretation
could have possible. However, a direct calculation of the dipole moment derivative with respect to the 
normal co-ordinates can serve the purpose. For, one makes the use of APT analysis that 
permits the interpretation of IR intensities 
by means of mean atomic charges ($\bar{p_{\alpha}}$), the mass-weighted square effective charges 
($\frac{\chi^{2}_{\alpha}}{m_{\alpha}}$) , where m$_{\alpha}$ and $\chi_{\alpha}$ 
are the mass and effective charge of each atom $\alpha$ of ClCCCN, since all these invariants are related to each other 
by the well known relation called as the G-sum rule \cite{guide43} given by: 
$\sum_{k} I_{k} = 974.9 \sum_{\alpha} \frac{3\chi^{2}_{\alpha}}{m_{\alpha}} - \Omega $, 
where $\Omega$ is the rotational correction term in km/mole arises from $P_{\rho}\beta$. 
Therefore, it permit a direct calculation of atomic polar charges.\\ 
\begin{table*}[htb]
\caption{ Comparison of the normal mode frequencie$s^{a}$ (in $cm^{-1}$) of
ClCCCN calculated at all the three different levels of theory with their corresponding experimental values. Here A, B, and 
C stands for 6-311++g(d,p), aug-cc-PVDZ and aug-cc-PVTZ basis sets respectively.}
\label{table:V}
\newcommand{\m}{\hphantom{$-$}}
\newcommand{\cc}[1]{\multicolumn{1}{c}{#1}}
\renewcommand{\tabcolsep}{0.05pc} 
\renewcommand{\arraystretch}{1.4} 
\begin{tabular}{@{}cccccccccccccccccc}
\hline
\hline
& \cc{$\nu_{i}$} & \cc{Species} & \cc{Motion$^{b}$} & & \cc{RHF} &            & & \cc{B3LYP}  & &      & \cc{B3PW91} &   & \cc{MP$_{2}$}&\cc{Expt.$^{c}$} &\cc{Expt.$^{d}$}\\
&                &              &                   &A  &B       &C           &A &B            &C &A      &B             &C   &A              &     &                  \\
\hline
\\
&1& $\Sigma^{+}$ &$ -C\equiv$N  st.         &\m2652.9&\m265.9&\m2644.3&\m2384.4&\m2382.2&\m2375.9&\m2397.3 &\m2395.8&\m2386.7&\m2263.7&\m2297 &\m2283\\
&2&$\Sigma^{+}$ &-C$\equiv$C  st.          &\m2477.0 &\m2478.9&\m2474.6&\m2248.3&\m2243.8&\m2249.2 &\m2256.8 &\m2253.1&\m2256.0&\m2087.6 &\m2194&\m2195\\
&3& $\Sigma^{+}$ &$\equiv$C-C$\equiv$ st.   &\m1152.6&\m1154.4&\m1148.4&\m1116.5&\m1118.2&\m1115.1 &\m1132.6 &\m1133.8&\m1129.5&\m1112.6 &\m1093&\m1099\\
&4&$\Sigma^{+}$ &$\equiv$C-Cl st.          &\m563.4  &\m564.4&\m563.0&\m540.1 &\m540.8 &\m541.3&\m549.1  &\m549.6&\m549.3&\m538.2 &\m527  &\m528\\
&5&$\Pi       $ &$\equiv$C-C$\equiv$N  bd. &\m582.8 &\m580.9&\m615.6&\m507.3 &\m496.2   &\m535.2&\m511.9  &\m497.5&\m536.8&\m412.6 &\m483  &\m482\\
&6&$\Pi       $ &$\equiv$C-C$\equiv$C- bd. &\m379.1  &\m397.8&\m423.5&\m268.2 &\m311.6   &\m361.6&\m268.0  &\m313.6&\m364.8&\m129.9 &\m333  &\m332\\
&7&$\Pi       $ &-C$\equiv$C-Cl  bd.       &\m150.0  &\m142.4&\m150.5&\m135.7 &\m121.5  &\m137.0&\m135.4  &\m123.4&\m137.5&\m\underline{434.2} &\m129 &\m140 \\
\\
\hline
\end{tabular}\\[2pt]
$^{a}$ Unscaled harmonic frequencies. \\
$^{b}$ st. = stretching; bd. = bending. \\
$^{c}$ Ref. \cite{guide25,guide26}. Normal mode frequencies observed in vapour phase. \\
$^{d}$ Ref. \cite{guide25,guide26}. Normal mode frequencies were observed in benzene. 
For comparison, the values of the normal modes 
are calculated at the B3LYP/aug-cc-pVDZ level of theory in solution (benzene) to be 2374.2, 2239.4, 1121.6, 
542.10, 500.7, 313.1 and 121.9 cm$^{-1}$ for $\nu_{1}$.... and $\nu_{7}$ respectively. \\ 
\end{table*}
\begin{figure}
\resizebox{8.0cm}{8.5cm}
{\includegraphics{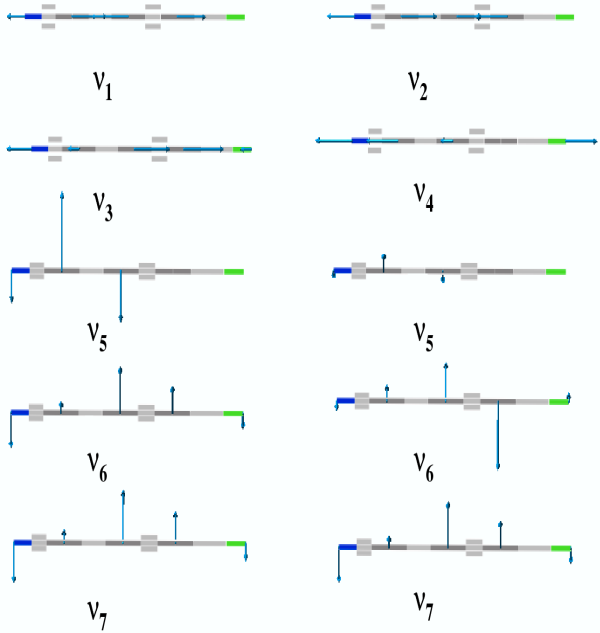}}    
\caption{Normal modes (summarized in Table 5) of ClCCCN. Atomic numbering is shown in Figure 1. 
The modes corresponding to $\nu_{5}-\nu_{7}$ are repeated twice since these are doubly 
degenerate. The actual directions of vibrations of the doubly degenerate bending modes were out of (shown as up) 
and into (shown as down) the paper which being modified here for a clear visualization of these modes.}
\label{fig3}                                 
\end{figure}
\begin{table*}[htb]
\caption{ Comparison of the IR intensities (in Km/mol) of
ClCCCN calculated at the HF and DFT levels with corresponding experimental values. Here A, B, and C stands for 6-311++g(d,p), aug-cc-PVDZ and aug-cc-PVTZ basis sets respectively.}
\label{table:V}
\newcommand{\m}{\hphantom{$-$}}
\newcommand{\cc}[1]{\multicolumn{1}{c}{#1}}
\renewcommand{\tabcolsep}{0.3pc} 
\renewcommand{\arraystretch}{1.4} 
\begin{tabular}{@{}ccccccccccccccccc}
\hline
\hline
& \cc{$I_{i}$}  &   & \cc{RHF} &          & & \cc{B3LYP}  & &      & \cc{B3PW91} &   &\cc{MP$_{2}$}& \cc{$A^{a}_{expt.} $} & \cc{$A^{b}_{expt.} $}\\
&               &A  &B       &C           &A &B            &C &A      &B             &C   &A              &             &   \\
\hline
&1 &\m195.05 &\m190.18 &\m185.27 &\m215.44 &\m200.78 &\m194.59 &\m219.57  &\m201.19 &\m195.93 &\m155.84 &\m vs &\m vs\\
&2 &\m38.88   &\m33.34  &\m40.98  &\m27.41  &\m19.47  &\m26.30   &\m31.02   &\m21.94  &\m28.92  &\m18.84 &\m w &\m w \\
&3 &\m16.26  &\m15.10  &\m14.47  &\m25.40  &\m22.60  &\m21.74   &\m27.12   &\m23.74  &\m22.60  &\m16.14 &\m s  &\m s\\
&4 &\m5.49    &\m4.73  &\m4.77  &\m8.15   &\m7.15     &\m7.01  &\m8.23    &\m7.17  &\m7.01  &\m8.11 &\m  m &\m m \\
&5 &\m9.20    &\m10.70 &\m9.62  &\m5.98   &\m7.91     &\m6.93  &\m5.44    &\m7.94  &\m6.84  &\m8.68 &\m s   &\m s\\
&6 &\m4.54    &\m7.37  &\m10.16 &\m5.22   &\m4.26     &\m7.37  &\m4.88    &\m4.03  &\m7.08  &\m2.36 &\m m   &\m m\\
&7 &\m4.87    &\m5.01  &\m4.91  &\m3.90   &\m3.78     &\m3.54  &\m3.86    &\m3.61  &\m3.38  &\m1.26 &\m vw  &\m vw\\
\hline
&Sum &\m274.29 &\m266.43 &\m270.18 &\m299.0 &\m265.95 &\m267.48 &\m291.50 &\m269.48 &\m271.76 &\m211.23 \\
\hline
\end{tabular}\\[2pt]
$^{a}$ Ref. \cite{guide25,guide26}. vs $=$ very strong, s $=$ strong, m $=$ medium strong; vw $=$ very weak; w $=$ weak. \\
$^{b}$ Ref. \cite{guide25,guide26}. For comparison, the IR intensities are 
calculated at the B3LYP/aug-cc-pVDZ level of theory 
in solution (benzene) to be 428.7, 21.8, 33.6, 10.2, 12.6, 4.2 and 5.4 cm$^{-1}$ for A$_{1}$... and A$_{7}$ respectively.  
\end{table*}
$~~~$The mean atomic charges as well as the other rotational invariants derived
from the atomic polar tensors (APTs) using equations (1) to (5) 
for all the five atoms of the ClCCCN are summarized in Table 4. The 
atomic polar charges $\bar{p_{\alpha}}$ represent the redistribution of the electric charge density around
each atom in such a way that their sum over all atoms equal to to zero in the molecular bonding environment of ClCCCN at 
its optimized equilibrium geometry. 
An important difference between APT charges and the Mullikan charges \cite{guide44,guide45}
is that the basis-set dependence of the
former arises only from the fact that the basis set can be incomplete; hence, as the basis set approaches
completeness, the APT charges approach a well-defined limit \cite{guide46}.
A test with ClCCCN, shown in Table 4, clearly indicates
that GAPTs are almost invariant with respect to the basis sets.
The polar tensor elements are calculated to be all negative for chlorine and nitrogen
atoms as expected since these are the electronegative elements, contrast to
the carbon atom (C$_{4}$) attached to it through a triple bond character which is highly positive, 
resulting the mean dipole
moment of nitrogen as negative and that of carbon as positive, each of which can be interpreted as carrying 
slightly more than -0.35 e and +0.35 e. This, in turn, results in an unequal distribution of charges among the two carbon
atoms, with some negative charge located on the carbon atom ($C_{3}$) bonded to $C_{4}$ and
an almost equal amount of positive charge on the other carbon atom $C_{2}$, leading to an interpretation
that the carbon atoms $C_{3}$ and $C_{2}$ carries atomic charges slightly more than +0.25e and -0.25e
respectively. The average values of atomic polar charges calculated by all the methods are -0.146e, 0.396e, -0.251e, 
0.366e and -0.365e having standard deviations $\pm$0.013e, $\pm$0.001e, $\pm$0.0.016e, $\pm$0.032e and $\pm$0.024e 
for Cl, C, C, C, N of ClCCCN respectively which indicates an overestimation of standard 
deviations of C$_{4}$ and N that arises mainly due to p$_{zz}$ component of the atomic polar tensors of C and N. It is noted 
worthy that all other invariant quantities are within 0.1e whereas a larger variation in the value of anisotropy
parameter $\beta$ as well as in the value of undeformability of charge ($\bar{P}/\beta$) on the atom $\alpha$ of ClCCCN (as shown in Table 4) has been noticed. \\
$~~~$The Mulliken atomic charges, on the other hand, of chlorine and carbon attached each other by means of 
a single bond are 
predicted to be positive in conjunction with 6-311++g(d,p) basis set at all the methods, whereas these are predicted to be 
negative in conjunction with aug-cc-pVNZ (N $=$ D, T) basis sets. Similarly, the charge of carbon atom attached to 
the nitrogen atom through a triple bond character is predicted to be negative irrespective of the methods and basis 
sets used. However, it successfully predicted a real sign to the nitrogen atom as expected. 
As can be seen, variations of 
these charges with respect to all the methods as a function of basis set choice is rather large, variations being -0.208e to 0.51e, -0.898e to 0.101e, 
1.155e to 1.985e, -0.387e to -2.154e and -0.117e to 0.454e for Cl, C, C, C and N of ClCCCN respectively, whereas APT 
charges are less variable ireespective of the methods and basis sets used. 
A comparison of APT 
charges with the Mulliken atomic charges indicates that the APT charges are highly reliable than the Mulliken atomic charges 
and the Mulliken population analysis of atomic charge densities of ClCCCN does not reflect the physical and chemical 
characteristics of this system under consideration. Finally, a comparison of the Mulliken atomic charges with 
the effective charges ($\chi$), shown in Table 4, reveals an increase of Chlorine(Cl), $C_{2}$, $C_{4}$ and $N_{5}$ APT 
charges as $^{M}q$ diminishes, which leads to an interpretation that $\chi$ is 
dominated by the stretching intensities \cite{guide47}. \\
\section{Conclusion}
All the spectroscopic constants were calculated at the restricted 
HF-SCF as well as DFT (B3LYP, B3PW91) levels in conjunction 
with a variety of basis sets. Satisfactory agreements between the B3LYP/6-311++g(d,p) and experimental values of 
rotational constants were found for ClCCCN. Over-all, $r_{Cl...N}$, bond distance is 
a good approximation to the over-all bond distance at the correlated B3LYP/6-311++g(d,p) level of theory. The values of 
chlorine and nitrogen NQCCs have been obtained at the B1LYP/TZV(3df,3p) 
and B3LYP/6-311++g(df,pd) levels respectively on the r$_{s}$ structure 
as well as on the B3PW91/6-311++g(d,p) optimized geometry are well within the scale length of experimental uncertainty. 
Enlarging basis sets size could improve the calculation accuracy and satisfactorily reproduced the 
experimental bending mode frequencies without being scaled with uniform scaling factors, though the stretching frequencies 
are a bit off. The rotational invariants have been satisfactorily explained at these levels of theory. 
The stationary point found on the potential energy surface of ClCCCN 
calculated at the MP$_{2}$/6-311++g(d,p) level of theory is not a true minimum rather a saddle point of order two 
and thus scan of the potential energy surface should be emphasized in oder to determine the nature of the saddle point 
and verify any possible existence of conical intersection between the ground and first excited state.  

{\it Acknowledgments.}{\small The author would like to thank Dr. Nikhil Guchhit of University of Culcutta, 
for kindly allowing to use Gaussian 03 suite program package and Dr. P. R. Bangal for helpful discussions 
in connection with Gaussian. The author would like to acknowledge 
Dr. R. L. A. Haiduke for helpful discussions.}


\end{document}